\documentclass[12pt]{article}
\usepackage{amsfonts}
\usepackage{amssymb}
\usepackage{graphicx}
\usepackage{amsmath}

\input{tcilatex}
\input{tcilatex}
\begin{document}

\title{On the first solution of a long standing problem: uniqueness of the\\
phaseless quantum inverse scattering problem in 3-d}
\author{Michael V. Klibanov}
\date{Department of Mathematics and Statistics, University of North Carolina
at Charlotte, Charlotte, NC 28223, USA\\
E-mail: mklibanv@uncc.edu}
\maketitle

\begin{abstract}
After publishing his recent paper in SIAM J. Appl. Math, 74, 392-410, 2014

the author has realized that actually he has addressed in that paper, for

the first time, a long standing open question being unaware about this. This

question is about the uniqueness of a 3-d inverse scattering problem without

the phase information. Thus, it makes sense in the current paper to

explicitly make the latter statement and to formulate corresponding

uniqueness theorems.
\end{abstract}

\graphicspath{
{FIGURES/}
 {pics/}}

\section{Introduction}

In quantum scattering one is measuring the differential scattering cross
section. In the frequency domain, this means that the modulus of the
scattering complex valued wave field is measured. However, the phase is not
measured. On the other hand, the entire theory of the inverse quantum
scattering problem is based on the assumption that both modulus and phase
are measured outside of a compact support of a scatterer. The latter was
noticed in Chapter 10 of the well known book of K. Chadan and P.C. Sabatier
published in 1977 \cite{CS}. In particular, Chadan and Sabatier state in
introduction to Chapter 10 \textquotedblleft \emph{In typical situations,
the scattering phase remains deeply involved in the formulas. Therefore,
mathematical ways of constructing the scattering amplitude from the cross
sections will be of interest for years}".

Recently the author has proved, for the first time, uniqueness theorems for
the phaseless inverse scattering problem (PISP) in the 3-d case \cite{KSIAP}%
. However, only after the paper \cite{KSIAP} was published, the author has
realized that \cite{KSIAP} is the first publication where the long standing
problem posed in \cite{CS} is addressed, at least partially. Thus, the goal
of the current paper is to draw attention to this fact. We formulate here
two of our four theorems of \cite{KSIAP} and briefly outline ideas of their
proofs. The idea of \cite{KSIAP} was extended in \cite{Kacoustics} to the
case of a 3-d phaseless inverse problem for the acoustic equation. While 
\cite{KSIAP} is about the 3-d case, the 1-d phaseless case was first
considered by the author and Sacks in \cite{KlS}.

\section{Statements of Problems}

Below $C^{s+\alpha }$ are H\"{o}lder spaces, where $s\geq 0$ is an integer
and $\alpha \in \left( 0,1\right) .$ Let $\Omega ,G\subset \mathbb{R}^{3}$
be two bounded domains, $\Omega \subset G$ and $S=\partial G$ be a piecewise
continuous boundary of $G$. We assume that%
\begin{equation}
dist\left( S,\partial \Omega \right) \geq 2\varepsilon ,  \label{2.1}
\end{equation}%
where\ the number $\varepsilon >0$ and $dist\left( S,\partial \Omega \right) 
$ denotes the Hausdorff distance. Let the potential $q\left( x\right) ,x\in 
\mathbb{R}^{3}$ be a real valued function such that 
\begin{equation}
q\left( x\right) \in C^{m+\alpha }\left( \mathbb{R}^{3}\right) ,q\left(
x\right) =0\text{ for }x\in \mathbb{R}^{3}\diagdown \Omega ,q\left( x\right)
\geq 0,\forall x\in \Omega .  \label{2.2}
\end{equation}%
Here either $m=2$ or $m=4,$ as will specified later. Let $x_{0}\in S$ be the
position of the source. As the forward problem, we consider the following 
\begin{eqnarray}
\Delta _{x}u+k^{2}u-q\left( x\right) u &=&-\delta \left( x-x_{0}\right)
,x\in \mathbb{R}^{3},  \label{1.4} \\
u\left( x,x_{0},k\right) &=&O\left( \frac{1}{\left\vert x-x_{0}\right\vert }%
\right) ,\left\vert x\right\vert \rightarrow \infty ,  \label{150} \\
\sum\limits_{j=1}^{3}\frac{x_{j}-x_{j,0}}{\left\vert x-x_{0}\right\vert }%
\partial _{x_{j}}u\left( x,x_{0},k\right) -iku\left( x,x_{0},k\right)
&=&o\left( \frac{1}{\left\vert x-x_{0}\right\vert }\right) ,\left\vert
x\right\vert \rightarrow \infty .  \label{1.5}
\end{eqnarray}%
Here the frequency $k\in \mathbb{R}$ and conditions (\ref{150}), (\ref{1.5})
are valid for every fixed source position $x_{0}.$ We now refer to works of
Vainberg \cite{V1,V} as well as to a classical result about elliptic PDEs of
the book of Gilbarg and Trudinger \cite{GT}. Theorem 3.3 of the paper \cite%
{V1}, Theorem 6 of Chapter 9 of the book \cite{V} in combination with
Theorem 6.17 of \cite{GT} guarantee that for each pair $\left(
k,x_{0}\right) \in \mathbb{R\times R}^{3}$ there exists such a unique
solution $u\left( x,x_{0},k\right) $ of the problem (\ref{1.4}), (\ref{150}%
), (\ref{1.5}) that can be represented in the form%
\begin{equation}
u\left( x,x_{0},k\right) =u_{0}\left( x,x_{0},k\right) +u_{sc}\left(
x,x_{0},k\right) ,  \label{1.6}
\end{equation}%
\begin{equation}
u_{0}=\frac{\exp \left( ik\left\vert x-x_{0}\right\vert \right) }{4\pi
\left\vert x-x_{0}\right\vert },\text{ }u_{sc}\in C^{m+2+\alpha }\left(
\left\{ \left\vert x-x_{0}\right\vert \geq \eta \right\} \right) ,\forall
\eta >0.  \label{1.7}
\end{equation}%
Here $u_{0}$ is the incident spherical wave and $u_{sc}$ the scattered wave.

\textbf{Phaseless Inverse Scattering Problem 1 (PISP1)}. \emph{Suppose that
the function }$q\left( x\right) $\emph{\ satisfies conditions (\ref{2.2}),
where }$m=2$\emph{. Assume that the following function }$f_{1}\left(
x,x_{0},k\right) $\emph{\ is known }%
\begin{equation}
f_{1}\left( x,x_{0},k\right) =\left\vert u\left( x,x_{0},k\right)
\right\vert ,\forall x_{0}\in S,\forall x\in \left\{ \left\vert
x-x_{0}\right\vert <\varepsilon \right\} ,x\neq x_{0},\forall k\in \left(
a,b\right) ,  \label{1.70}
\end{equation}%
\emph{where }$\left( a,b\right) \subset \mathbb{R}$\emph{\ is an arbitrary
interval. Determine the function }$q\left( x\right) $\emph{\ for }$x\in
\Omega .$

This inverse problem is over-determined.\ Indeed, the function $q\left(
x\right) $ depends on three variables, whereas the function $f_{1}\left(
x,x_{0},k\right) $ depends on six variables. However, it is well known that
uniqueness theorems for inverse scattering problems in 3-d in the case when
the $\delta -$function is considered as the source, are known only for the
over-determined case, even if both modulus and phase of the scattering field
are given outside of the scatterer, see, e.g. Chapter 6 of the book of
Isakov \cite{Is}.

Consider now the non over-determined case of the data resulting from a
single measurement. In this case uniqueness theorems for coefficient inverse
problems in $n-d$, $n\geq 2$ are known only if the $\delta -$function is
replaced with a regular function which is non-zero in the entire domain of
interest $\overline{\Omega }$ where the coefficient is unknown. All these
theorems were proven using the method, which was originated in the work of
Bukhgeim and Klibanov \cite{BukhKlib}, also see the recent survey of the
author \cite{Klib5}.\ This method is based on Carleman estimates. Thus, we
replace the $\delta -$function in (\ref{1.4}) with the function $g\left(
x\right) $ such that 
\begin{equation}
g\in C^{4}\left( \mathbb{R}^{3}\right) ,g\left( x\right) =0\text{ in }%
\mathbb{R}^{3}\diagdown G_{1},g\left( x\right) \neq 0\text{ in }\overline{G},
\label{1.10}
\end{equation}%
where $G_{1}\subset \mathbb{R}^{3}$ is a bounded domain such that $\Omega
\subset G\subset G_{1},\partial G\cap \partial G_{1}=\varnothing .$ As it
was mentioned in \cite{KSIAP,Klib5}, the function $g\left( x\right) $ can
be, for example an approximation for the $\delta -$function via a narrow
Gaussian-like function, which is equivalent to the $\delta -$function from
the Physics standpoint. As the forward problem, we consider the following%
\begin{eqnarray}
\Delta w+k^{2}w-q\left( x\right) w &=&-g\left( x\right) ,x\in \mathbb{R}^{3},
\label{1.12} \\
w\left( x,k\right) &=&O\left( \frac{1}{\left\vert x\right\vert }\right)
,\left\vert x\right\vert \rightarrow \infty ,  \label{1.120} \\
\sum\limits_{j=1}^{3}\frac{x_{j}}{\left\vert x\right\vert }\partial
_{x_{j}}w\left( x,k\right) -ikw\left( x,k\right) &=&o\left( \frac{1}{%
\left\vert x\right\vert }\right) ,\left\vert x\right\vert \rightarrow \infty
.  \label{1.13}
\end{eqnarray}%
The same results of \cite{GT,V1,V} as ones mentioned above guarantee that
for each $k\in \mathbb{R}$ there exists unique solution $v\left( x,k\right)
\in C^{5+\alpha }\left( \mathbb{R}^{3}\right) ,\forall \alpha \in \left(
0,1\right) $ of the problem (\ref{1.12}), (\ref{1.120}), (\ref{1.13}).

\textbf{Phaseless Inverse Scattering Problem 2 (PISP2)}. \emph{Assume that
in (\ref{2.2}) }$m=4$\emph{, the function }$q\left( x\right) $\emph{\
satisfying conditions (\ref{2.2}) is unknown for }$x\in \Omega $\emph{\ and
known for }$x\in \mathbb{R}^{3}\diagdown \Omega $ \emph{and that the
function }$g\left( x\right) $\emph{\ satisfies conditions (\ref{1.10}).\
Determine the function }$q\left( x\right) $\emph{\ for }$x\in \Omega $\emph{%
\ assuming that the following function }$f_{2}\left( x,k\right) $\emph{\ is
known }%
\begin{equation}
f_{2}\left( x,k\right) =\left\vert v\left( x,k\right) \right\vert ,\forall
x\in S,\forall k\in \left( a,b\right) .  \label{1.14}
\end{equation}

We now outline the main difficulty in addressing each of above inverse
problems. We consider only the PISP1, since this difficulty is similar for
the PISP2. For a fixed pair $\left( x,x_{0}\right) $ denote $p\left(
k\right) =u\left( x,x_{0},k\right) ,k\in \mathbb{R}$. Thus, the modulus $%
\left\vert p\left( k\right) \right\vert $ is known for all $k\in \left(
a,b\right) $. For each number $\gamma >0$ denote 
\begin{equation*}
\mathbb{C}_{\gamma }=\left\{ z\in \mathbb{C}:\func{Im}z>-\gamma \right\} ,%
\text{ }\mathbb{C}_{+}=\left\{ z\in \mathbb{C}:\func{Im}z>0\right\} .
\end{equation*}%
The function $p\left( k\right) $ admits the analytic continuation from the
real line $\mathbb{R}$ in the half-plane $\mathbb{C}_{\gamma }$ for a
certain number $\gamma >0$ (see details below)$.$ For each $z\in \mathbb{C}$
let $\overline{z}$ be its complex conjugate. Since 
\begin{equation}
\left\vert p\left( k\right) \right\vert ^{2}=p\left( k\right) \overline{p}%
\left( k\right) ,\forall k\in \mathbb{R},  \label{100}
\end{equation}%
then the function $\left\vert p\left( k\right) \right\vert ^{2}$ is analytic
for $k\in \mathbb{R}$ as the function of the real variable $k$. Hence, the
modulus $\left\vert p\left( k\right) \right\vert $ is known for all $k\in 
\mathbb{R}.$ Let $\left\{ a_{j}\right\} _{j=1}^{n}\subset \mathbb{C}_{+}$ be
all zeros of the function $p\left( k\right) $ in the upper half plane. Here
and below each zero is counted as many times as its multiplicity is. Using
an analog of classical Blaschke products \cite{Col}, consider the function $%
p_{a}\left( k\right) $ , 
\begin{equation}
\widetilde{p}\left( k\right) =p\left( k\right) \dprod\limits_{j=1}^{n}\frac{%
k-\overline{a}_{j}}{k-a_{j}}.  \label{1.190}
\end{equation}%
The function $\widetilde{p}\left( k\right) $ is analytic in $\mathbb{C}%
_{\gamma }$ and $\left\vert p_{a}\left( k\right) \right\vert =\left\vert
p\left( k\right) \right\vert ,\forall k\in \mathbb{R}$. Therefore, the
central question is about finding of complex zeros. To do this, one needs to
figure out how to combine the knowledge of $\left\vert p\left( k\right)
\right\vert $ for $k\in \mathbb{R}$ with a linkage between the function $%
p\left( k\right) $ and the original forward problem (\ref{1.4})-(\ref{1.5}).

\section{Uniqueness Theorems and Main Ideas of Proofs}

\textbf{Theorem 1}. \emph{The PISP1 has at most one solution.}

\textbf{Theorem 2}. \emph{The PISP2 has at most one solution.}

We now outline main ideas of proofs of these theorems. We start from Theorem
1. Consider the following hyperbolic Cauchy problem%
\begin{eqnarray}
v_{tt} &=&\Delta _{x}v-q\left( x\right) v,\left( x,t\right) \in \mathbb{R}%
^{3}\times \left( 0,\infty \right) ,  \label{3.1} \\
v\left( x,0\right) &=&0,v_{t}\left( x,0\right) =\delta \left( x-x_{0}\right)
.  \label{3.2}
\end{eqnarray}%
Let $\Phi \subset \mathbb{R}^{3}$ be an arbitrary bounded domain. Using
lemma 6 of Chapter 10 of \cite{V} as well as Remark 3 after that lemma, one
can proof that for each fixed source position $x_{0}\in \mathbb{R}^{3}$ the
function $v\left( x,x_{0},t\right) $ decays exponentially with respect to $%
t\rightarrow \infty $ uniformly for all $x\in \Phi ,x\neq x_{0}$ together
with its derivatives involved in (\ref{3.1}). One can derive from the latter
that 
\begin{equation}
u\left( x,x_{0},k\right) =\dint\limits_{0}^{\infty }v\left( x,x_{0},t\right)
e^{ikt}dt=\mathcal{F}\left( v\right) ,\forall x,x_{0}\in \mathbb{R}%
^{3},x\neq x_{0},\forall k\in \mathbb{R},  \label{3.3}
\end{equation}%
where $\mathcal{F}$ is the operator of the Fourier transform (\ref{3.3}).
Hence, for every pair $x,x_{0}\in \mathbb{R}^{3},x\neq x_{0}$ there exists a
number $\gamma =\gamma \left( x,x_{0}\right) >0$ such that the function $%
u\left( x,x_{0},k\right) $ can be analytically continued with respect to $k$
in the half-plane $\mathbb{C}_{\gamma }.$ Next, using (\ref{3.1})-(\ref{3.3}%
), we derive the behavior of the function $u\left( x,x_{0},k\right) $ for $%
\left\vert k\right\vert \rightarrow \infty ,k\in \mathbb{C}_{+}$ and
conclude that this function has at most finite number of zeros in $\mathbb{C}%
_{+},$ for each above pair $x,x_{0}.$

Now the turn is for the most difficult part of the proof: we should prove
that these zeros can be uniquely determined using the function $f_{1}\left(
x,x_{0},k\right) $ in (\ref{1.70}). Assume that there exist two potentials $%
q_{1}\left( x\right) $ and $q_{2}\left( x\right) $ satisfying conditions (%
\ref{2.2}) with $m=2$ and producing the same function $f_{1}\left(
x,x_{0},k\right) $. Let $u_{1}\left( x,x_{0},k\right) $ and $u_{2}\left(
x,x_{0},k\right) $ be solutions of the problem (\ref{1.4})- (\ref{1.5}) with
functions $q_{1}\left( x\right) $ and $q_{2}\left( x\right) $ respectively.
Fix two arbitrary points $x_{0}\in S$ and $x\in \left\{ y:\left\vert
y-x_{0}\right\vert <\varepsilon \right\} ,x\neq x_{0}.$ Consider
corresponding functions $h_{1}\left( k\right) =u_{1}\left( x,x_{0},k\right) $
and $h_{2}\left( k\right) =u_{2}\left( x,x_{0},k\right) .$ Then both these
functions are analytic in $\mathbb{C}_{\gamma }$ and $\left\vert h_{1}\left(
k\right) \right\vert =\left\vert h_{2}\left( k\right) \right\vert ,\forall
k\in \left( a,b\right) .$ First, using (\ref{100}), we show that 
\begin{equation}
\left\vert h_{1}\left( k\right) \right\vert =\left\vert h_{2}\left( k\right)
\right\vert ,\forall k\in \mathbb{R}  \label{3.4}
\end{equation}%
and that real zeros of functions $h_{1}\left( k\right) $ and $h_{2}\left(
k\right) $ coincide.

As to the more difficult case of complex zeros in $\mathbb{C}_{+},$ we
consider analogs Blaschke products (\ref{1.190}).\ Let $\left\{
a_{j}\right\} _{j=1}^{n_{1}}\subset \mathbb{C}_{+}$ and $\left\{
b_{j}\right\} _{j=1}^{n_{2}}\subset \mathbb{C}_{+}$ be all zeros of
functions $h_{1}\left( k\right) $ and $h_{2}\left( k\right) $ respectively
in the upper half plane. Then (\ref{3.4}) implies that 
\begin{equation}
h_{1}\left( k\right) +h_{1}\left( k\right) \left( \prod\limits_{j=1}^{m}%
\frac{k-b_{j}}{k-\overline{b}_{j}}-1\right) =h_{2}\left( k\right)
+h_{2}\left( k\right) \left( \prod\limits_{j=1}^{n}\frac{k-a_{j}}{k-%
\overline{a}_{j}}-1\right) .  \label{3.5}
\end{equation}%
Next, we use the partial fraction expansion in (\ref{3.5}) and the following
formula which can be proved via a straightforward computation for every
integer $s\geq 1$ 
\begin{equation}
\mathcal{F}^{-1}\left( \frac{1}{\left( k-\overline{d}\right) ^{s}}\right)
=H\left( t\right) \frac{\left( -i\right) ^{s}}{\left( s-1\right) !}%
t^{s-1}\exp \left( -i\overline{d}t\right) ,\text{ }\forall d\in \mathbb{C}%
_{+},  \label{3.6}
\end{equation}%
where $H\left( t\right) $ is the Heaviside function, $H\left( t\right) =1$
for $t>0$ and $H\left( t\right) =0$ for $t<0.$ Let $v_{j}\left(
x,x_{0},t\right) =\mathcal{F}^{-1}\left( h_{j}\right) ,j=1,2.$ Since $%
q_{1}\left( x\right) =q_{2}\left( x\right) =0$ outside of the domain $\Omega 
$, $x\in \left\{ \left\vert x-x_{0}\right\vert <\varepsilon \right\} $ and
by (\ref{2.1}) $\left\{ \left\vert x-x_{0}\right\vert <\varepsilon \right\}
\cap \overline{\Omega }=\varnothing ,$ then (\ref{3.1}) and (\ref{3.2})
imply that $v_{1}\left( x,x_{0},t\right) =v_{2}\left( x,x_{0},t\right) $ for 
$t\in \left( 0,\left\vert x-x_{0}\right\vert +\varepsilon \right) .$ Hence,
applying the operator $\mathcal{F}^{-1}$ to both sides of (\ref{3.5}), using
(\ref{3.6}) and the convolution theorem for the Fourier transform, we obtain
that there exists a function $\lambda \left( t\right) $ which depends on
above zeros $\left\{ a_{j}\right\} _{j=1}^{n_{1}},\left\{ b_{j}\right\}
_{j=1}^{n_{2}}$ such that it satisfies a homogeneous Volterra integral
equation of the second kind for sufficiently small values of $t>0$. Hence, $%
\lambda \left( t\right) =0$ for sufficiently small $t>0$. Next, since $%
\lambda \left( t\right) $ is an analytic function of the real variable $t$
for $t>0,$ then $\lambda \left( t\right) =0$ for all $t>0.$ This leads to
the conclusion that $\left\{ a_{j}\right\} _{j=1}^{n_{1}}=\left\{
b_{j}\right\} _{j=1}^{n_{2}}$ and $n_{1}=n_{2}.$ This, in turn implies that
even if the function $f_{1}\left( x,x_{0},k\right) $ in (\ref{1.70}) is
known for only a single pair $x_{0}\in S,x\in \left\{ \left\vert
x-x_{0}\right\vert <\varepsilon \right\} ,x\neq x_{0}$ and for $k\in \left(
a,b\right) ,$ then values of $u\left( x,x_{0},k\right) $ are still uniquely
determined for the same pair $x,x_{0}$ and for all $k\in \mathbb{R}.$

The final step is to prove that the potential $q\left( x\right) $ is
uniquely determined. To do this, we show that 
\begin{equation*}
\dint\limits_{L\left( x,x_{0}\right) }\left( q_{1}-q_{2}\right) \left(
x\right) ds=0,\text{ }\forall x,x_{0}\in S,x\neq x_{0}.
\end{equation*}%
Finally the uniqueness of the Radon transform implies that $q_{1}\left(
x\right) \equiv q_{2}\left( x\right) .$

The proof of Theorem 2 is similar with only one difference. Namely, on the
final step of the proof we use theorem 3.2 of \cite{Klib5} instead of the
uniqueness of the Radon transform.

\begin{center}
\textbf{Acknowledgments}
\end{center}

This research was supported by US Army Research Laboratory and US Army
Research Office grant W911NF-11-1-0399. The author is grateful to P.E.
Sacks, A. T. Synyavsky and B.R. Vainberg for their valuable comments.


\begin{thebibliography}{99}
\bibitem{BukhKlib} A.L. Bukhgeim and M.V. Klibanov, Uniqueness in the large
of a class of multidimensional inverse problems, \emph{Soviet Math. Doklady}%
, 17, 244-247, 1981.

\bibitem{CS} K. Chadan and P.C. Sabatier, \emph{Inverse Problems in Quantum
Scattering Theory}, Springer-Verlag, New York, 1977.

\bibitem{Col} P. Colwell, \emph{Blaschke Products: Bounded Analytic Functions%
}, University of Michigan Press, Ann Arbor, 1985.

\bibitem{GT} D.\ Gilbarg and N.S.\ Trudinger, \emph{Elliptic Partial
Differential Equations of Second Order}, Springer,\ New York, 1984.

\bibitem{Is} V. Isakov, \emph{Inverse Problems for Partial Differential
Equations}, Second Edition, Springer, New York, 2006.

\bibitem{KSIAP} M.V. Klibanov, Phaseless inverse scattering problems in
three dimensions, \emph{SIAM J. Appl.\ Math}., 74, 392-410, 2014.

\bibitem{Kacoustics} M.V. Klibanov, Uniqueness of two phaseless
non-overdetermined inverse acoustics problems in 3-d, accepted for
publication in \emph{Applicable Analysis}, available online of this journal
http://dx.doi.org/10.1080/00036811.2013.818136.

\bibitem{KlS} M.V. Klibanov and P.E. Sacks, Phaseless inverse scattering and
the phase problem in optics, \emph{J. Math.\ Physics}, 33, 3813-3821, 1992.

\bibitem{Klib5} M.~V.~Klibanov, {C}arleman estimates for global uniqueness,
stability and numerical methods for coefficient inverse problems, \emph{J.
Inverse and Ill-Posed Problems}, 21, 477-560, 2013.

\bibitem{V1} B.R. Vainberg, Principles of radiation, limiting absorption and
limiting amplitude in the general theory of partial differential equations, 
\emph{Russian Math. Surveys}, 21, 115-193, 1966.

\bibitem{V} B.R. Vainberg, \emph{Asymptotic Methods in Equations of
Mathematical Physics}, Gordon and Breach Science Publishers, New York, 1989.
\end{thebibliography}
\end{document}